# Giant and Linear Magnetoresistance in Liquid Metals at Ambient Temperature

Xiaolin Wang[1,2]*, Feixiang Xiang[1], David Cortie[1], Zengji Yue[1], Zhi Li[1], Zhidong Zhang[3] and Lina Sang[1]

[1] Institute for Superconducting and Electronic Materials, Australian Institute for Innovative Materials, University of Wollongong, Wollongong, NSW 2500, Australia.

[2] ARC Centre for Excellence in Future Low Energy Electronics Technologies, University of Wollongong, Wollongong, NSW 2500, Australia.

[3] Magnetism and Magnetic Materials Division, Shenyang Materials Science National Laboratory, Institute of Metal Research, Chinese Academy of Sciences, 72 Wenhua Road, Shenyang 110016, P. R. China

*Email: xiaolin@uow.edu.au

**Abstract**: **Disorder-induced magnetoresistance has been reported in a range of solid metals and semiconductors, however, the underlying physical mechanism is still under debate because it is difficult to experimentally control. Liquid metals, due to lack of long-range order, offers an ideal model system where many forms of disorder can be deactivated by freezing the liquid. Here we report non-saturating magnetoresistance discovered in the liquid state of three metals: Ga, Ga-In-Sn and Bi-Pb-Sn-In alloys. The giant magnetoresistance appears above the respective melting points and has a maximum of 2500% at 14 Tesla. The reduced diamagnetism in the liquid state implies that a short-mean free path of the electron, induced by the spatial distribution of the liquid structure, is a key**

**factor. A potential technological merit of this liquidtronic magnetoresistance is that it naturally operates at higher temperatures.**

**Summary:** Giant magnetoresistance is observed in the liquid phase of three metals, with a linear dependency on magnetic field.

**Main Text:** From supercomputers to social media, the phenomenon of magnetoresistance (MR) has revolutionized society by enabling information storage on an unprecedented scale. The operational principle behind MR sensors is the large change in resistance ($\rho$) in response to a magnetic field (B) with a definition described as the ratio MR = [$\rho$(B) - $\rho$(0)]%/[$\rho$(0)]. The MR ratio is sensitive to details in the electronic structure of materials[1, 2], geometry[3, 4] and material disorder[5]. On the one extreme, MR in non-magnetic solid metals occurs at the level of a few percent which is too low to be practically useful. At the other limit, colossal MR over thousands of percent is known in rare earth magnetite's within a narrow temperature range near the metal-insulator transition[6], and recently larger effects have been discovered in exotic semimetals e.g. $WTe_2$ which include a linear momentum-energy relationship in their band structures[7, 8]. High MR over a broad temperature range is a desirable feature in candidate materials, however, it is very rare. While it is possible to increase the MR ratio by applying stronger magnetic fields, this approach is limited in any semi-classical electron system (i.e. most metals and semiconductors) because the dependency is parabolic at low fields, (MR $\propto$ $B^2$) but it ultimately reaches a saturation point at high fields. A significant discovery in the late 1990's was a class of disordered materials such as $Ag_2Te$[9, 10] yielding giant MR (GMR) that keeps increasing with fields above 10 Tesla and never saturates. Indeed, similar phenomena were reported even in the early days of condensed matter physics and have been considered an enigma for decades[11]. Electrons moving in materials with an open Fermi surface, such as noble metals, will never form a closed orbital under a magnetic field and therefore yield unsaturated MR[12]. However, it is now known that non-saturating MR occurs in a wide range of 3D[8, 13] and 2D materials[14, 15] with closed Fermi surfaces. To explain the broad range of observations, and its correlation with disorder, recent theories have proven that the minimum criteria for this remarkably useful effect is that the local conductivity must be inhomogeneous on a length scale larger than the mean free-path of the electron[10, 16]. In this scenario, simple classical arguments lead to a situation where the Lorentz force introduces current drift counter to the external potential[10, 17]. One distinctive feature of disorder-induced MR is that the parabolic-dependency at low fields generally transitions to a linear-dependency at high fields. Unlike quantum MR[1] or the special-case of polycrystalline metals with open Fermi surfaces[18], the classical mechanism for non-saturating MR does not depend closely on the details of the electronic structure, and there is no strict requirement for single-occupied Landau levels or a compensated electron and hole band. Recently, it has been argued that this effect is the ubiquitous in inhomogeneous conductive materials[10], and a method was developed to assign a universality class characterized by the functional form of the quadratic-to-linear transition[5]. Given these theoretical advances, and with the benefit of hindsight, it seems surprising that the MR characteristics remains almost unknown for an entire class of naturally disordered matter: the liquid state.

In this article, we report giant linear MR in three liquid metals at high fields, with no signature of saturation up to 14 Tesla. The results are compelling because MR has been scarcely studied in the flowing states of warm condensed matter as common liquids do not possess sufficient electronic conductivity. Liquid metals are the rare exception, providing an ideal model system with spatially-disordered ionic cores that are heavily screened by a nearly-free electron gas. Our results indicate that charge-disorder is a natural byproduct of the structural disorder in the liquid state, and occurs with a characteristic length-scale well-matched to the electron transport length-scale in order to yield MR. In addition to opening a new field for MR research, our work unveils insights into the structure of the liquid state more generally, because the MR appears to imply the existence of medium range order that is difficult to characterize using other methods.

The key experimental result of our work is that a significantly enhanced MR appears in the liquid phase of the metals, which then vanishes upon cooling to the solid state (Fig. 1a). The measurements are based on standard techniques using a Quantum Design Physical Property Measurement System (PPMS) adapted to investigate transport phenomena in liquid metals (See Methods Section), with the caveat that the motion of the liquid drop needs to be constrained for reliable measurements across the melting point. The liquid metals should also be thick enough to minimize any influence from oxide surfaces which naturally form in an ambient environment. While there is only a small change in resistivity for $B = 0$, the liquid-solid transition exhibits a large change in the resistivity at higher fields ($B > 1$ T). The moderate increase in resistivity measured at 0 T is consistent with past zero-field resistivity measurements performed since the 1950's[19, 20] that culminated in a well-established model relating the average conductivity with the average static liquid structure factor $S(\mathbf{q})$ [20, 21]. On the other hand, the transport under high magnetic fields ($B > 1$ T) have not been studied deeply experimentally or theoretically. An enigmatic feature of the MR in the molten state is the linear dependency on field that appears at higher fields (Fig. 1b) producing giant MR with no sign of saturation up to 14 Tesla. This behavior is quite distinct from that of the frozen solid which only shows MR at the level of a few percent with a weak parabolic dependency (Fig. 1b). The results can be rationalized using recent theoretical models which have shown that, using the four-probe transverse-field geometry (Fig. S1), the resistivity becomes related to the root-mean-squared variance of the spatially varying conductivity, whereby the Lorentz force induces currents counter to the electric potential [14,15,16]. In contrast, for zero-field measurements, only the spatially-averaged conductivity is detected.

Further measurements in two other liquids metals strongly suggest that there is a universal mechanism operating in the liquid state. We studied two other molten metals: 99.999% Gallium and a Bi-Sn-In-Sb eutectic alloy. Despite the very different underlying chemistry, both materials showed a large, linear MR (Fig. 2a,b) above their respective melting points. We hope that this work will stimulate many more studies to assess whether this feature is truly universal in liquid metals, however, the initial results are immensely optimistic. One clear difference between the three liquid metals we studied, is that each displays a different region where quadratic dependency survives (shaded purple in the figure), up to a field defined as $B_q$ which ranges from low (0.2 T) to higher fields (1 T).

To shed light on the underlying phenomena, we renormalize the data by dividing B by $B_q$ and MR by the its value at $B_q$ using the procedure outlined in Ref.[5]. $B_q$ is taken as the field where value of the least-squared fitting parameter ($\chi^2$) diverges when fitting MR to a model $x^2$ function. This normalization is necessary, because although the MR for the three metals is qualitatively similar, each spans a different quadratic width. After replotting the MR in rescaled units of $B/B_q$ and $MR/MR(B_q)$, it is striking that all three liquid metals tend to collapse towards the same universal curve (Fig. 3). Indeed, as expected, the experimental function falls within the region of allowed quadratic-to-linear functions bounded in the natural units by the functions $y = x^2$ and $y = 2x-1$ (for $x>1$). The values for $B_q$ are 0.2 T, 0.5 T and 1.1 T for Ga-In-Sn, Ga and Bi-Sn-In-Sb alloy respectively. After these measurements, we also became aware of a single group who performed experiments in liquid mercury and also detected a giant, enhanced liquid-state MR, with a quadratic dependency to at least 5 T[22]. It is currently unknown whether the liquid-Hg-MR displays linear dependency, since it appears that $B_q$ may be much higher there, and different measurement geometry may also play a role[22, 23]. Therefore, the unique unsaturated linear MR in liquid metals can be phenomenologically explained by the RRN model, although other atomic-scale factors unique to the liquid state should also be considered.

In a single solid crystal, a saturated MR will generally be measured if all electron and hole orbits at the Fermi surface are closed. A non-saturated MR can be observed, however, if some orbits are open, which was well proved in MR experiments on noble metals with certain magnetic field orientations[12]. There is an exception to this principle. In a compensated two-band metal (one electron band and one hole band) with zero net charge carrier, the MR will not saturate and increase as $B^2$ with increasing field[12]. This non-saturating MR has been recently observed in Weyl semimetal $WTe_2$[7]. However, all these MR measurements require very low temperature and very pure samples with large mean free path. In liquid metals, there are significant natural imperfection in structure and electronic structure due to lack of long range order which reduce the electrons' mean free path. However, atoms in liquid metals are not totally randomly distributed. Unlike a gas where the atoms are far apart, the distance between two atoms in a liquid cannot be significantly shorter than the size of atom and also cannot exceed more than a few times the atom diameter. X-ray and neutron scattering experiments indicate that there is structure correlation in radial direction only over short distances (i.e. angstroms). Thus although the powerful Bloch theorem cannot be used in liquid metals, a pseudo-potential method was introduced on the basis of a near-free electron model to explain the zero-magnetic-field conductivity of liquid metals based on their average radial structure, which matched well with experiment results[24, 25]. In this theory, the Fermi surface is assumed to be spherical. By extension, the randomness in structure will induce multiple scattering of electrons when an electric field and a strong magnetic field are applied. The multiple random scattering will hinder the formation of closed cyclotron orbitals, thus favoring a non-saturating MR[26].

To induce a *linear* non-saturating MR in disordered solids, a further criterion is that the local conductivity is modulated spatially on a 'macroscopic' length-scale larger than the mean free path of the electron[16, 17]. This argument holds equally in liquid metals which are described by a nearly free electron model as established in Ziman's theory[24, 25]. The 'macroscopic' criterion is automatically satisfied in the liquid for several reasons. Firstly, the electron mean free path is short in liquid metals owing to lack of long range order (16, 17). The theoretical mean free path for liquid metal Ga, In, Sn and Pb is 17.4 Å, 16.6 Å, 9.9 Å and 5.4 Å, respectively[24], and the experimental results indicate somewhat shorter paths[27]. Secondly, the variation in conductivity in liquid metals occurs on a spectrum of different time-scales and spatial-scales tied to the local structure and dynamics. While the average static structure factor S(q) is well-defined over several angstrom, it also conceals short and medium range order on the scale of nanometers[28] together with long-range density fluctuations[21], and local point-defects from short-lived cage-type dynamics that modify the disorder potential. From a structural perspective, a substantial body of research shows that short-range icosahedral order dominates in liquids, however, liquid metals sometimes contain regions with degrees of face-centred-cubic or hcp order[28]. For liquid gallium, studies suggest the local structure shares motifs with the high pressure solid-crystal phases of Ga-II/Ga-III[29] as supported by experiment[30]. More generally, molecular dynamic (MD) simulations have identified medium range-order in liquid metals, such as ''Bergman triacontahedron'' packing that extends out to the fourth shell (i.e. more a nanometre)[28]. First principle calculations have been used to connect the conductivity and local structure for various liquid metals, and predictably, the conductivity is found to vary strongly for different local trial structures, even when the overall average configurations produce the correct average $S(q)$[31, 32]. *Ab initio* calculations based on Kubo-Greenwood formalism indicate that local variations in time and space generate different conductivities in liquid metals even when large up supercells ~ 1 nm$^3$ are used[32]. Although the authors of those works did not frame their arguments in this way, the latter findings broadly support the concept that the electronic environment in liquid metals varies over a relatively long length-scale relative to the electron mean free path, thereby modifying conductivity. Indeed, early work already established the long-range oscillatory potential from the ions in a liquid metal extends to over a nanometre[33] in analogy to the screened Friedel oscillations surrounding charged defects in solid metals. It appears that the length-scale of the structural disorder, charge disorder and electron transport conspire to yield optimal conditions for MR in liquid metals.

A full theoretical treatment of liquid metal MR for realistic local structures under high magnetic field does not yet exist in the literature, which is peculiar given that strong magnetic fields universally modify electron transport leading to emergence of critical effects including Shubnikov de Haas oscillations and MR. The effect of charge disorder was tentatively discussed in Ziman's early theory, however, past work has placed emphasis on calculating the average zero-field conductivity and this parameter is insensitive to local fluctuations[14]. Within the well-established Ziman's theory for liquid metals, the average conductivity is directly related to the average static liquid structure factor S(**q**)[24, 25]. The success of this approach to predict the zero-field resistance is reasonable, and rests on the conflating all structural details into an average interference function which modifies the electrons treated as planewaves under a pseudopotential[20, 21]. The notion of spatially-average conductivity, however, breaks down once large magnetic fields are applied[14], whereupon local inhomogeneity becomes increasingly important[34]. With the Ziman framework the relative insensitivity of the zero-field resistivity to the solid-liquid phase transition is predictable because these systems are governed by a nearly-free electron (NFE) model due to heavy screening of the disorder ion cores, reducing the sensitivity to the precise lattice structure. Weak electron-electron interactions also yield an effective mass ~ 1 in both the liquid and solid state. The obvious discrepancy between the solid and the liquid is that the latter has no long-range translation symmetry in real space, and any local structural variations, which are "averaged out" in the angular-averaging underlying the static S(**q**) picture, will result in modulations of the local conductivity. If the conductivity is modulated by local variations of liquids' radial distribution function, as in the spirit of Ziman's theory, and this occurs on a length-scale greater than the electron mean free path, then the basic prerequisites for the mechanism of semi-classical disorder-induced MR are satisfied.

Using magnetometery measurements, we can also exclude one alternative possibility highlighted in a recent theoretical proposal[35], termed the guiding center theory. In this model, disorder-induced linear MR arises from the interaction of the cyclotron guiding center motion with a smooth random potential arising from material disorder with a restricted spatial scale compared to the cyclotron orbital length. In this dynamic picture, the electron undergoes rapid field-induced cyclotron orbits around a slow-drifting guiding center. For the key dynamic to survive, an electron is required to complete many cyclotron motions before being scattered. If this proposal is valid to liquid metals, an enhancement of Landau orbital diamagnetism should be observed due to the increased electron cyclotron motions. Figure 4 plots the measured magnetization as a function of temperature on Ga and Ga-In-Tin samples. The results agree fairly well with previously published susceptibility measurements for liquid gallium[36]. In contrast to the guiding center theory, the diamagnetism is much weaker at liquid state. The weaker diamagnetism is a natural consequence of the highly disordered structure in the liquid metal state which perturbs cyclotron orbits via scattering, and therefore decreases the orbital contribution to diamagnetism. From the free electron model, the nominal cyclotron radius of a free electron at 10 T is as large as 11.4 nm, whereas experimentally the mean free path for liquid gallium may be less than a single nanometer[27]. While this rules out the premises of the guiding-center theory, it also supports the concept that the electron short-mean free path is a key factor that leads to linear MR. A corollary of this is that any quantum theory of linear MR in liquid metals cannot be fully valid, and semi-classical factors are probably dominant.

In conclusion, our results demonstrate giant MR is ubiquitous in liquid metals. New avenues are now open to test fundamental theories of magnetotransport in the presence of extremely high density disorder. Technologically, the large liquid-state MR above room temperature is attractive, and by exploiting this together with the thermal memory characteristics of the solid-liquid transition, liquidtronic GMR could become paramount for flexible magnetoelectronics, sensors and phase-change memory that operates even into the high temperature regime. From a fundamental perspective, the large linear MR sheds new light on the contemporary debate surrounding the origin of linear giant MR which has been also widely observed within Dirac materials such as topological insulators[8]. The liquid metal results provides an incontrovertible counter argument, because we observed giant linear MR in a system with no well-defined band structure, arising from classical effects, therefore complementing recent observations of linear MR in a GaAs 2D electron gas with parabolic dispersion[3, 15]. We have shown at least three liquid metals fall within a new class of linear MR, however it is unclear why this class is distinct from all previous classes of disordered induced MR known in solids. On general grounds, a unique feature of three-dimensional liquids is that the fourth dimension (time) can also play a strong role owing to the rapid ionic motions. From the perspective of the electrons, most of the liquid dynamics, such as the diffusion processes, are very slow (< 1 THz) and thus effectively appear "frozen" to the electronic sub-system. However, certain ultra-fast femtosecond ionic dynamics are reported in liquid metals, resulting from a rare cage-type motion[37]. These ultra-fast motions, which break the Born Oppenheimer approximation, will result in a disorder field that shifts on the same time-scale as the electron motion, although the overall importance to the total disorder field is difficult to predict. A complete understanding of giant-liquid MR is therefore a truly fascinating theoretical and experimental challenge with implications for a promising new field of liquidtronics.

**Acknowledgments:**

This work is supported by Australian Research Council through an ARC Professorial Future Fellowship project (FT130100778, X.L.W) and the ARC Centre of Excellence in Future Low-

**Figures and Captions**

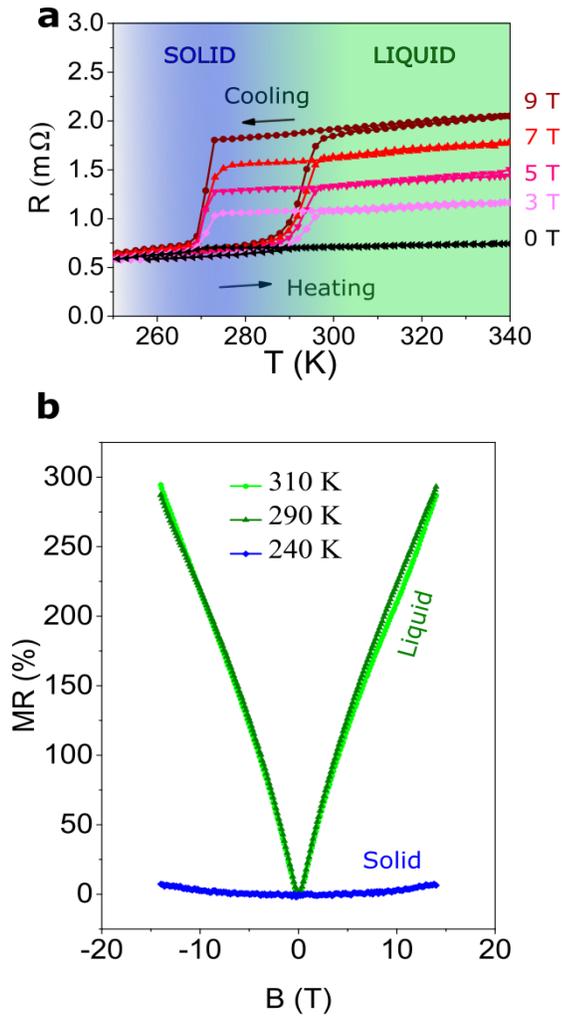

**Figure 1.** Magnetoresistance measurements in a Ga-In-Sn eutectic alloy above and below its melting point. a) The magnetoresistivity of the metal is strongly increased in the liquid state, and exhibits thermal hysteresis at the first-order transition. b) The liquid state is characterized by a linear, unsaturating high field.

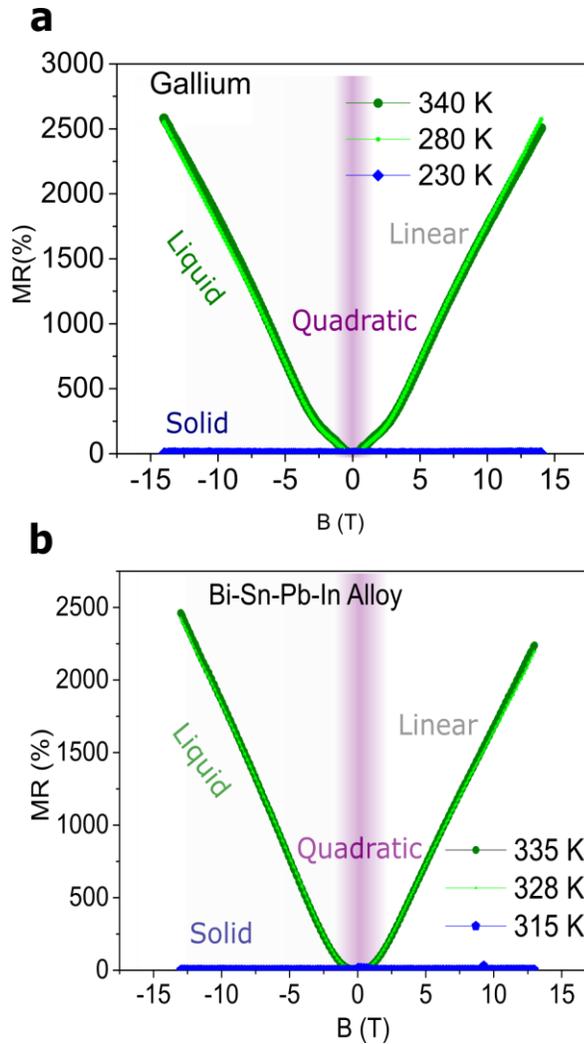

**Figure 2.** Magnetoresistance measurements in pure gallium and a Bi-Sn-Pb-In alloy showing a gradual transition from quadratic-to-linear dependency. a) The pure gallium exhibits giant magnetoresistance up to 2500% near room temperature. b) The average MR for the droplet of Bi-Sn-Pb-In is 2250% and exhibits quadratic dependency over a wider range of fields. The purple regions indicate the region where MR is quadratic.

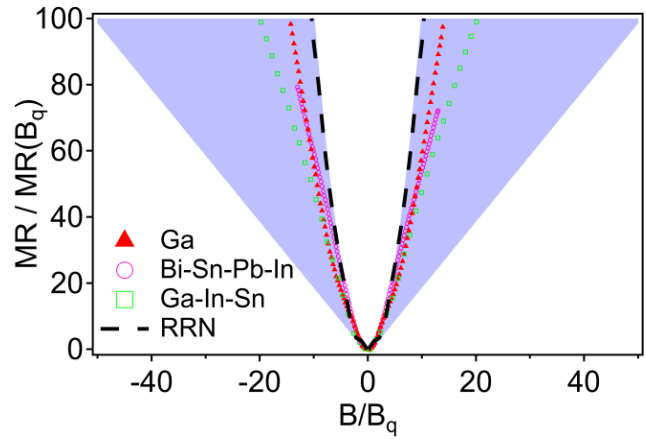

**Figure 3.** Rescaled MR data for the liquid state of EuGaIn, Ga and Bi-Sn-Pb-In in natural units. The experimental data fo three collapse onto a universal curve which is quite different from the random-resistor network (RRN) class reported for solid-state materials. The blue shaded region formed from the upper and lower bounds of the quadratic-linear transition.

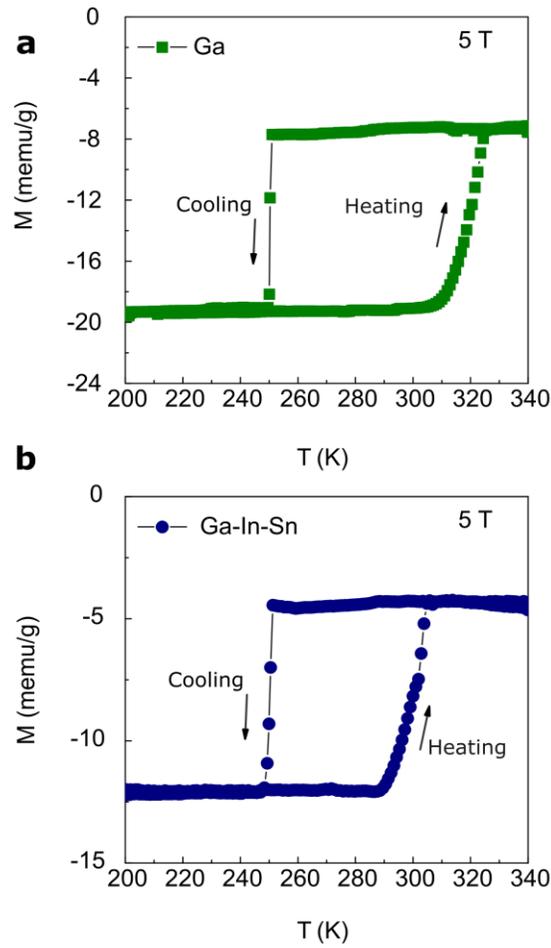

Figure 4. Plot of magnetization as a function of temperature. Both measurements on Ga (a) and Ga-In-Tin alloy (b) at 5 T show diamagnetic signal below melting points, indicating Landau diamagnetism dominating the liquid metal behavior. Above the melting point, there is sharp decreasing in diamagnetism, which is the consequence of presence of large amount of disorders that scatter electrons during incomplete cyclotron motions and quench the orbital contribution.